\documentstyle[preprint,eqsecnum,aps,prb,floats,epsfig]{revtex}
\begin{document}
\title{Nonlinear dynamic susceptibilities of interacting and noninteracting nanoparticle systems }
\author{P. J{\"o}nsson, T. Jonsson\cite{newadd}, J. L. Garc\'{\i}a-Palacios\cite{oldadd}, and P. Svedlindh}
\address{Department of Materials Science, Uppsala University\\
        Box 534, SE-751 21 Uppsala, Sweden}
\date{\today}
\maketitle

\begin{abstract}

The linear and cubic dynamic susceptibilities of solid dispersions of
nano-sized maghemite particles have been measured for three samples with a volume concentration of magnetic particles ranging from 0.3 \% to 17 \%, in order to study the effect of dipole-dipole interactions.
Significant differences between the dynamic response of the three samples are observed.
The dynamic susceptibilities of the most dilute sample compare reasonably well with an existing theory for the linear and cubic dynamic susceptibilities of an assembly of noninteracting, uniaxial magnetic particles. 
The nonlinear dynamic response of the most concentrated sample exhibits at low temperatures some of the features observed in a Ag(11 at.\% Mn) spin glass.
%
%

\end{abstract}


\section{Introduction}
\label{Intro}
The magnetic response of ensembles of single-domain magnetic particles has been in focus since the pioneering work of N\'{e}el and Brown. \cite{Neel,Brown} 
In the case of noninteracting particle systems, much theoretical work \cite{Coffey,Raikher1,Shliomis} have been devoted to the linear response.
Experimental results from magnetic relaxation and ac susceptibility measurements \cite{Wernsdorfer,Jonsson1} reasonably well support the now existing models.

The linear response of interacting magnetic particle systems is more involved. 
It has recently been shown that dipole-dipole interactions introduce collective behavior, as evidenced by the appearance of magnetic aging and a significantly broadened magnetic relaxation. \cite{Jonsson2,Jonsson3} 
Moreover, the dynamics of a magnetic particle system of monodispersive nature is indicative of critical slowing down at a finite temperature, \cite{Djurberg} implying the existence of a low temperature spin-glass-like phase. 

Much less work have focused on the nonlinear response of magnetic particle systems.
Bitoh and co-workers \cite{Bitoh,Shirane} studied experimentally the third harmonic of the ac magnetization to obtain information related to the cubic dynamic susceptibility.
However, the ac field used in this study was comparably large, \cite{Garcia} which makes it possible that the results for the third harmonic of the magnetization are contaminated with higher order susceptibility terms. 
In order to obtain theoretical expressions for the dynamic nonlinear susceptibility for systems without magnetic anisotropy, one can directly translate the expressions obtained for the dielectric relaxation. \cite{Coffey3}
The inclusion of magnetic anisotropy is more involved and it is only recently that this has been done for systems with the simplest uniaxial anisotropy.\cite{Garcia,Raikher2,Klik} 
%
%
Of special interest is the study of the cubic dynamic susceptibility by Raikher and Stepanov,\cite{Raikher2} who used the Fokker-Planck equation to obtain numerically exact results on the linear and cubic dynamic susceptibilities. 
They also suggested approximate analytical expressions for these quantities.

In the case of interacting magnetic particle systems, experimental work related to the nonlinear magnetic response are even more scarce. \cite{Fiorani,Jonsson4}
However, in a recent study it was shown that the {\it equilibrium} nonlinear response of an interacting magnetic particle system of monodispersive nature indicates that the cubic equilibrium susceptibility diverges at a finite temperature, \cite{Jonsson4} thus providing further evidence for a low temperature spin-glass-like phase. 
To the best of our knowledge, no experimental work has been reported for the {\it dynamic} higher order susceptibility terms of interacting particle systems.

In this paper, we report a study of the linear and cubic dynamic susceptibilities of a magnetic particle system consisting of nanosized maghemite particles.
The effects of interparticle interactions are investigated by studying three samples with different volume concentrations of magnetic particles.
The results for the most dilute sample are compared with  the expressions for the susceptibility proposed by Raikher and Stepanov.
The dynamic response of the most concentrated sample is compared with the dynamic response of a Ag(11at.\%Mn) spin glass.


\section{Experimental}
\label{Exp}

The experiments were performed on three samples consisting of nanosized maghemite ($\gamma$-Fe$_{2}$O$_{3}$) particles, with a typical particle diameter of 6~nm \cite{ferrofluids}  and almost spherical particle shape (observed in TEM analysis). \cite{Jiang}
The particles were suspended in a hydrocarbon oil and coated with a surfactant layer preventing the particles from agglomerating. 
Since the measurements were performed at low temperatures, the oil was frozen and the particles fixed randomly in the sample. 
Three samples with different volume concentration of particles, $\sim$0.3\%, 3\%, and 17\%, were used in order to study the effects of magnetic dipole-dipole interactions.
The lowest concentration should be sufficiently low to make the dipole-dipole interactions between the particles negligible.
Experiments were also performed on a Ag(11at.\%)Mn spin glass sample exhibiting long range spin-spin interactions of RKKY type.
.
The spin glass sample was prepared by melting pure Ag and Mn together at 1000~$^{\circ}$C in an evacuated and sealed silica tube.
After annealing the sample at 850~$^{\circ}$C for 72h it was quenched in water to room temperature. 

Two different experimental equipments have been used: 
(i) A commercial ac-susceptometer \cite{LakeShore} was used to measure the 1st and 3rd harmonics of the magnetization for different ac-field amplitudes in the range $100$ -- $2000$ A/m. 
(ii) A non-commercial low-field superconducting quantum interference device (SQUID) magnetometer \cite{SQUID}  was used to perform sensitive studies of the frequency dependence of the linear and cubic dynamic susceptibilities. 
Frequencies in range of 2 Hz to 200 Hz were used.


The magnetization, $M$, can be expanded with respect to an applied field $H$ as
\begin{equation}
M = \chi H + \chi_{3} H^{3} + \chi_{5} H^{5} + \cdots,
\end{equation}
where $\chi$ is the linear susceptibility and $\chi_{3}$ the cubic susceptibility. 
The dynamic susceptibility can be probed by applying an ac field, $H=h_{0}\cos(\omega t)$. 
The linear susceptibility is then obtained from the magnetization  measured at the fundamental frequency as $\chi(\omega)=M_\omega/h_0$ and the cubic dynamic susceptibility is obtained from the third harmonic of the magnetization as $\frac{1}{4}\chi_{3}(\omega)=M_{3\omega}/h_{0}^{3}$.
These equations are only valid if  the applied ac field is sufficiently small, so that contributions from higher order susceptibility terms in the expansion of $M_\omega$ and $M_{3\omega}$ are negligible.
For the measurements of the linear and cubic dynamic susceptibilities we used different ac-fields to ascertain that no mixing with higher order susceptibility terms occured.
%

\section{Theoretical background}
\label{Theo}

\subsection{Equilibrium susceptibilities}

The potential energy of a single-domain particle in an external magnetic field is the sum of the Zeeman energy and the anisotropy energy.
For a particle with uniaxial anisotropy the total magnetic potential is given by 
\begin{equation}
U = - D(\hat{e} \cdot \hat{n})^{2} - \mu_{0} M_{s} V(\hat{e} \cdot \vec{H}),
\label{PotE}
\end{equation}
where $D$ is the anisotropy energy barrier and $\vec{H}$ is the external filed, while $\hat{n}$ and $\hat{e}$ are the unit vectors along the anisotropy axis and magnetic moment, respectively.
In the case of volume anisotropy $D=KV$, where $V$ is the particle volume and $K$ is the volume anisotropy constant, and for surface anisotropy $D=K_{S}S$, where $S$ is the particle surface and $K_{S}$ is the surface anisotropy constant.


The linear and cubic {\it equilibrium} susceptibilities can be derived from the partition function associated with the magnetic potential.
For a mono-dispersive system with randomly distributed anisotropy axes, the linear and cubic equilibrium susceptibilities are given by\cite{Garcia,Raikher2}
\begin{equation}
\chi^{\rm eq}(T) =  \frac{\mu_{0} M_{s}^{2} V}{3 k_{B} T}, \qquad 
\chi_{3}^{\rm eq}(T) =  - \frac{\mu_{0}^{3} M_{s}^{4} V^{3}}{(k_{B} T)^{3}}
\frac{(1+2S_{2}^{2})}{45},
\label{susceq}
\end{equation}
respectively, where
\begin{equation}
S_{2}(\sigma) = Z_{0}^{-1}(\sigma) \int_{-1}^{1}{P_{2}(z)\exp(\sigma z^{2})dz}.
\label{S2}
\end{equation}
Here $P_{2}(z)=\frac{1}{2} (3z^{2}-1)$ is the second  Legendre polynomial, $z = (\hat{e} \cdot \hat{n})$,  $\sigma = D/k_{B}T$, and $Z_{0}=\int_{-1}^{1}\exp(\sigma z^{2})dz$ is the partition function in the absence of an applied magnetic field. 

\subsection{Dynamic susceptibilities}

Recently, Raikher and Stepanov \cite{Raikher2} studied theoretically the linear and cubic {\it dynamic} susceptibilities of noninteracting ensembles of single-domain particles with uniaxial anisotropy, solving numerically the Fokker-Planck equation in the overdamped case. 
It is difficult to derive analytical expressions for these quantities, but for a particle system with randomly distributed anisotropy axes, they suggested the following  simple formulae for the linear dynamic susceptibility,
\begin{equation}
{\chi(T,\omega)} = \chi^{\rm eq}
\left [ 
\frac{1+2S_{2}(\sigma)}{3(1 + i \omega \tau)}+\frac{2}{3}(1-S_{2}(\sigma)) 
\right].
\label{theo1}
\end{equation}  
This expression has been shown to be a good approximation to the exact linear dynamic susceptibility for frequencies below the ferromagnetic resonance frequency regime (see, for instance, Ref.\onlinecite{Garcia2}).
In the above expression $\tau$ is the relaxation time, for which various analytical expressions have been suggested (see, for instance, Ref.~\onlinecite{Cregg}). 
In the moderate to high-energy barrier case, the relaxation time is approximately given by an Arrhenius law as $\tau=\tau_{0}\exp(D/k_{B}T)$, where $\tau_{0}$ is a constant. 
Raikher and Stepanov also proposed an expression for the cubic dynamic susceptibility
\begin{equation}
{\chi_{3}(T,\omega)} = \chi_{3}^{\rm eq}
\frac{(1-i\omega \tau)}{(1+i\omega \tau)(1+3i\omega \tau)},
\label{theo3}
\end{equation}  
which they assert to be a good approximation of the numerically exact result in the low frequency regime.
From the above equations we can see that for $\omega \tau \ll 1$, e.g., for high temperatures, the dynamic susceptibilities reduce to the equilibrium susceptibilities Eq.~(\ref{susceq}).

\section{Results and Discussion}

The linear susceptibilities for the three maghemite samples measured at the frequency $\omega/2\pi = 125$ Hz and with an ac-field amplitude of 100~A/m are shown in Fig.~\ref{susc1_H}.
Dipole-dipole interactions shift the susceptibility peaks to higher temperatures, lower the magnitude of the peaks, but produce a higher equilibrium susceptibility.  
Figure~\ref{susc3_H}  shows the corresponding cubic susceptibilities measured with an ac-field amplitude of 400 A/m.
Similar to the case of the linear susceptibility, dipole-dipole interactions shift the susceptibility peaks to higher temperatures and reduce their magnitudes. 
The sample with the highest concentration of particles has a second positive peak at low temperatures in  both the real and imaginary components of the cubic susceptibility.

The results for the most dilute sample and the most concentrated sample will be discussed separately below.
For the most dilute sample, the linear and cubic susceptibilities are compared to the theoretical expressions for noninteracting particle systems discussed in Sec~\ref{Theo}.
The sample with the highest particle concentration has been reported to show collective behavior at low temperatures. \cite{Jonsson3} Its cubic dynamic susceptibility is therefore compared with $\chi_{3}$ obtained for the spin glass sample.

\subsection{Noninteracting particles}

To compare the measured linear and cubic susceptibilities with theoretical expressions, the polydispersivity of the particle system needs to be taken into account.
The conventional approach is to choose a trial volume distribution and determine its parameters by fitting theoretical curves to experimental data.
The log-normal distribution and the gamma distribution have been shown to work well for a variety of nanoparticle systems.
The log-normal distribution is given by
\begin{equation}
g(V) =
\frac{1}{\sqrt{2\pi}Vs}\exp\left(-\frac{\ln(V/V_{0})}{2s^{2}}\right),
\label{logn-dist}
\end{equation}
where $V_0$ is the median particle volume and $s$ is the logarithmic standard deviation.
The gamma distribution is given by
\begin{equation}
g(V) =
\frac{1}{\Gamma(1+\beta)V_{0}}\left(\frac{V}{V_{0}}\right)^{\beta}\exp(-V/V_{0}),
\label{gamma-dist}
\end{equation}
where $\Gamma(\cdot)$ is the gamma function and $V_0$ and $\beta$ are parameters related to the mean particle volume and standard deviation.

The susceptibility of a polydispersive system is calculated as $\chi_{\rm poly} = \int_{V}\chi_{\rm mono} g_{V} dV$, where $g_{V}$ is the volume weighted volume distribution and $\chi_{\rm mono}$ is either the linear or the cubic susceptibility given by Eq.~(\ref{theo1}) and Eq.~(\ref{theo3}), respectively.
In the theoretical model the anisotropy is assumed to have uniaxial symmetry.
In other work on $\gamma$-Fe$_{2}$O$_{3}$ nanoparticles, it has been shown that the dominating magnetic anisotropy is uniaxial and of surface type. \cite{Gazeau,Hasmonay}
In these studies particles exhibiting almost spherical particle shape and with diameters in the range 4.8 - 10 nm were investigated,\cite{Gazeau} which correspond well to the particles investigated in the present study.

\begin{figure}[htb]
\centerline{\epsfig{figure=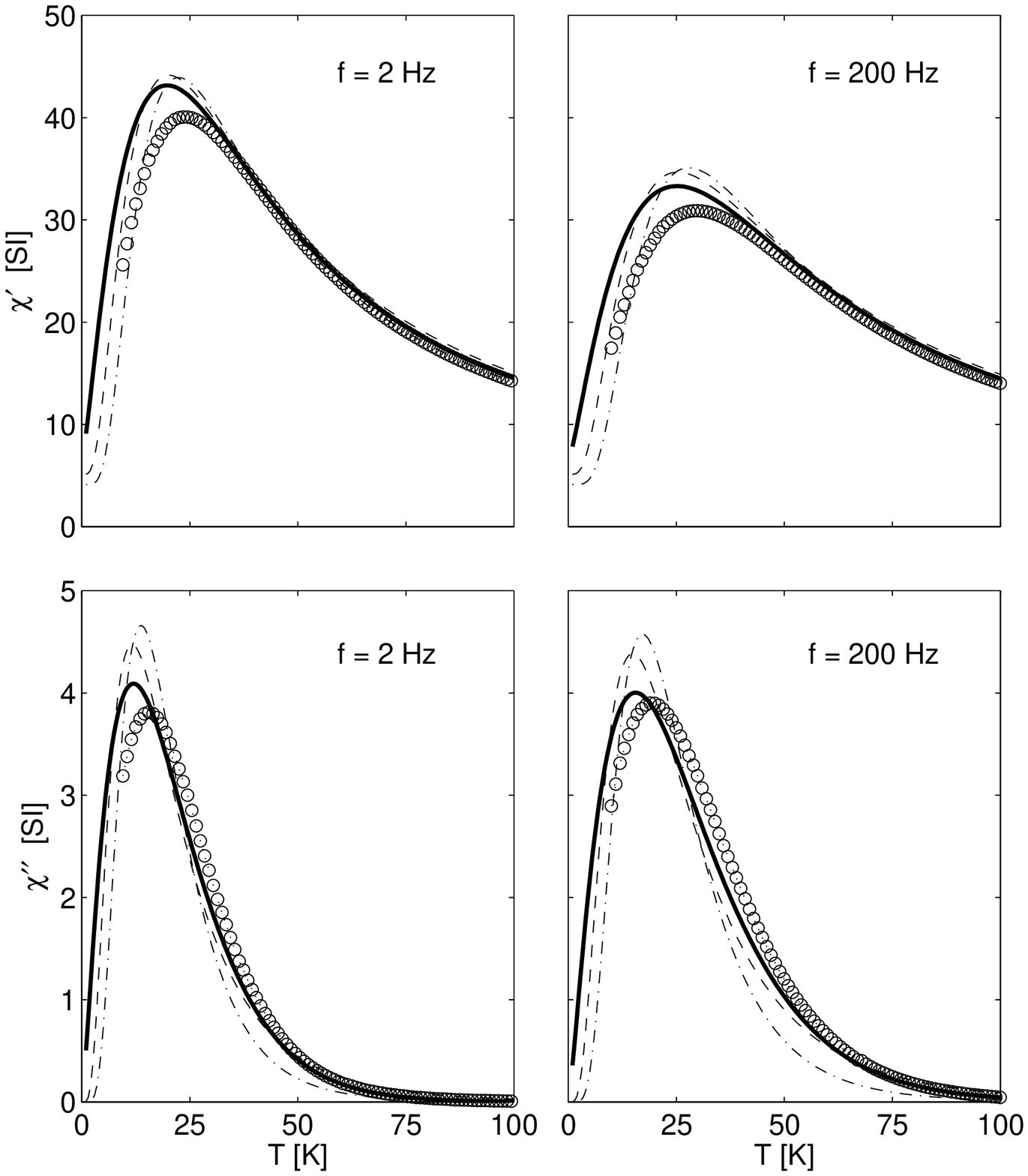,width=14truecm}}
\caption[]{The linear susceptibility vs. temperature for the sample with a volume concentration of 0.3\%. Solid lines represent the calculated susceptibility using volume anisotropy and the volume distribution taken as a gamma distribution, dashed lines show the corresponding result but with the volume distribution taken as a log-normal distribution and dashed-dotted lines represent the calculated susceptibility using surface anisotropy and the volume distribution taken as a log-normal distribution.}
\label{susc1}
\end{figure}

\begin{figure}[htb]
\centerline{\epsfig{figure=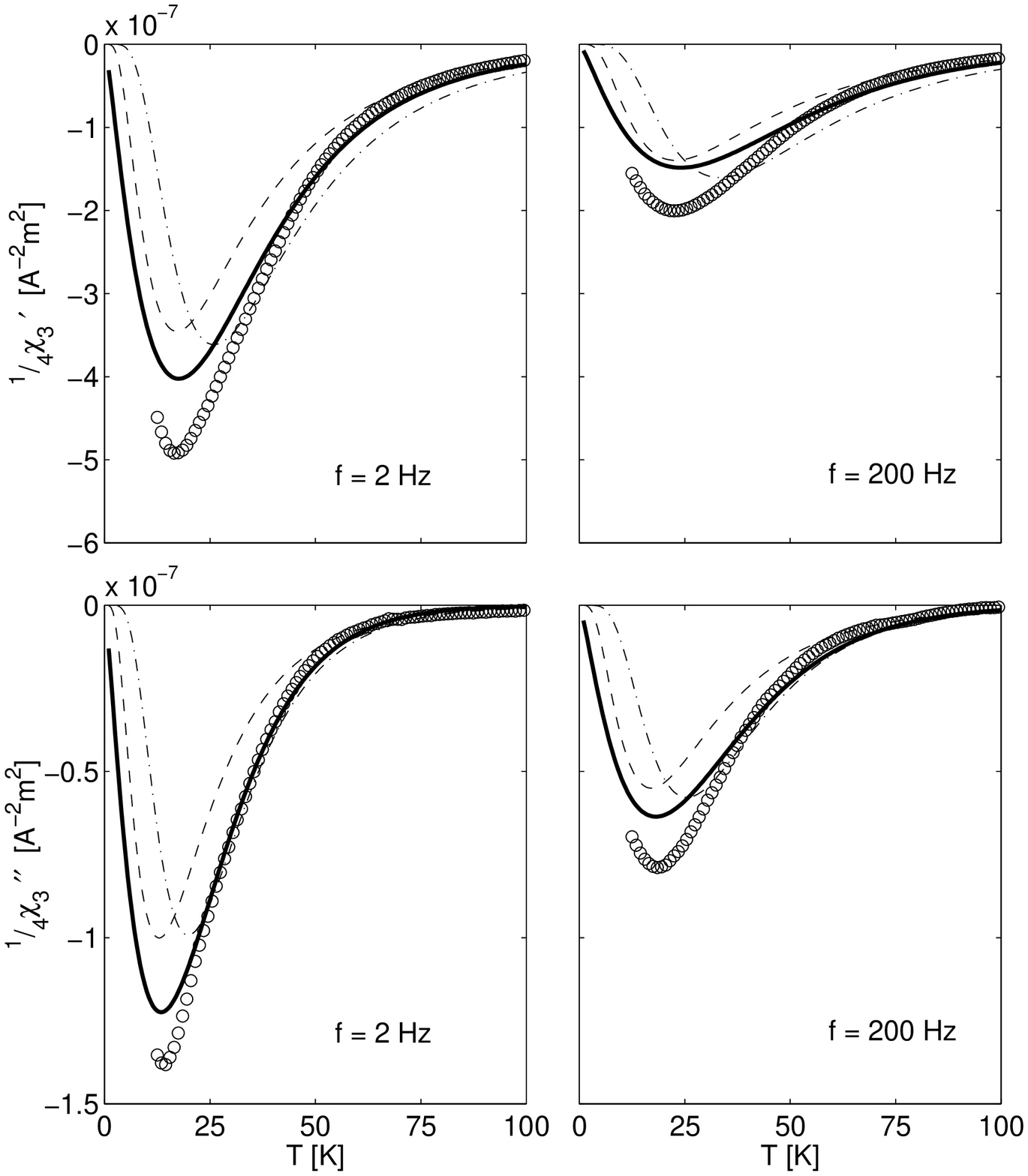,width=14truecm}}
\caption[]{The cubic susceptibility vs. temperature for the sample with a volume concentration of 0.3\%. Solid lines represent the calculated susceptibility using volume anisotropy and the volume distribution taken as a gamma distribution, dashed lines show the corresponding result but with the volume distribution taken as a log-normal distribution and dashed-dotted lines represent the calculated susceptibility using surface anisotropy and the volume distribution taken as a log-normal distribution.}
\label{susc3}
\end{figure}

The temperature dependence of the saturation magnetization at low temperatures has earlier been modeled as $M_s(T) = M_s(0) (1 - 1.8 \times 10^{-5} T^{3/2})$, with $M_s(0) = 4.2 \times 10^5$ A/m. \cite{Jonsson1}
The Nelder-Mead simplex method was used to perform nonlinear fitting to the measured linear and cubic susceptibilities for frequencies in the 2 Hz to 200 Hz range.
The fitting parameters were: the anisotropy constant, the preexponential factor in the expression for the relaxation time ($\tau_0$), and the two parameters from the volume distribution.
The parameters obtained from simultaneous fits to $\chi$ and $\chi_3$ are presented in Table~\ref{para}.
(We do not present the combination surface anisotropy and volume distribution modeled by a gamma distribution, since it gave unphysical fitting parameters.)
The anisotropy constant in the case of surface anisotropy $K_S = 2.3 \times 10^{-5}$ Jm$^{-2}$ is similar to the value $K_S = 2.7 \times 10^{-5}$ Jm$^{-2}$ obtained in Ref.~\onlinecite{Gazeau}. 
The different volume distributions give a typical particle diameter of $6\pm1$ nm in good agreement with the typical particle size of 6 nm observed in TEM studies.\cite{ferrofluids,Jiang}.
The results of the simultaneous fits to $\chi$ and $\chi_3$ are shown in Fig.~\ref{susc1} for the linear susceptibility and in Fig.~\ref{susc3} for the cubic susceptibility.
We can see that the best fit is obtained for a volume distribution modeled by a gamma distribution.
However,  the differences between the three cases are not sufficiently large to make any conclusions about the origin or the  uniaxial anisotropy or the functional form of the volume distribution in this sample.
We have also performed the fitting with the expression for the relaxation time proposed by Cregg {\it et al.} in Ref.~\onlinecite{Cregg}, but the change of the quality of the fit was negligible.

The discrepancies between the calculated and the measured dynamic cubic susceptibilities may have several possible origins. 
One obvious origin is that the models of the polydispersivity of the system used here are too simple and that the real situation is more involved. 
It may be that the symmetry of the magnetic anisotropy in the experimental system is different from uniaxial.
Such a difference need not originate from the magnetocrystalline anisotropy but can also be due to the geometrical shape of the individual particles. 
The lack of theoretical work where a magnetic anisotropy different from the uniaxial case is assumed makes it difficult to predict what effect the symmetry of the anisotropy will have on the nonlinear dynamic susceptibilities.
Another possible cause for the observed differences is linked to the details of the Fokker-Planck equation used in the theoretical work of Raikher and Stepanov, \cite{Raikher2} who studied the overdamped case (the damping constant, in the dynamical Landau Lifshitz equation from which the Fokker-Planck equation is derived, is assumed to fulfill  $\alpha \gg 1$). 
However, a more realistic value of the damping constant for magnetic nano-particles is $\alpha \lesssim 1$.\cite{Coffey2}
The form of the dynamic linear susceptibility curves is not very sensitive to the value of the damping constant (it essentially changes the value of $\tau_{0}$), but the same does not necessarily hold for the nonlinear response and a relatively more pronounced dependence of $\chi_{3}(\omega)$ on the damping parameter cannot, in principle, be ruled out.
Taking all these considerations into account, we can conclude that measurements of both the  linear and cubic susceptibilities potentially allow for extracting more detailed information concerning the intrinsic properties of real particle samples.

\subsection{Interacting particles}

\begin{figure}[htb]
\centerline{\epsfig{figure=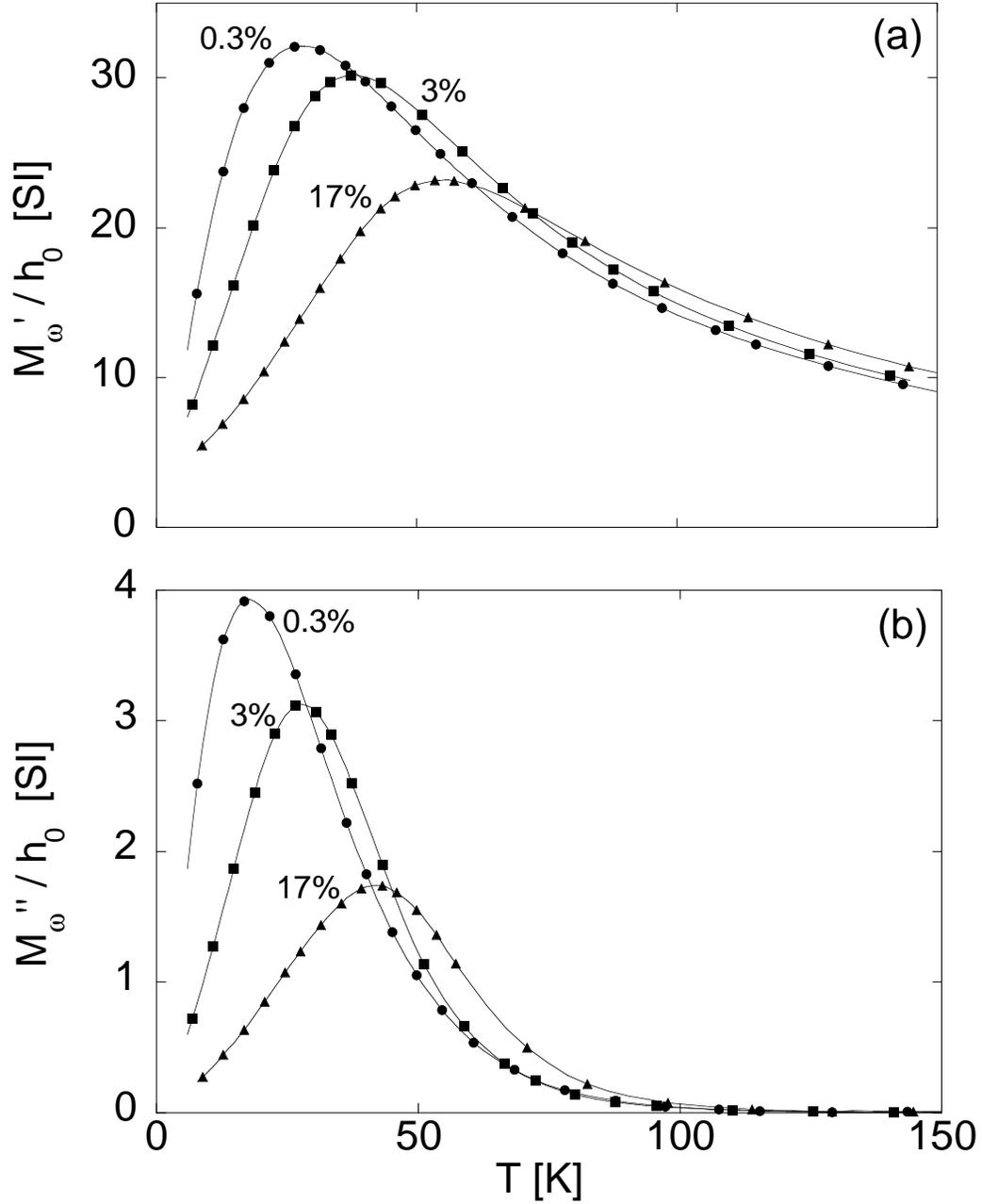,width=14truecm}}
\caption[]{The real (a) and the imaginary (b) parts of the linear susceptibility vs. temperature, for the three samples with different volume concentration of magnetic particles. $\omega/2\pi$ = 125 Hz and  $h_0$ = 100 A/m.
\label{susc1_H}
}
\end{figure}

\begin{figure}[htb]
\centerline{\epsfig{figure=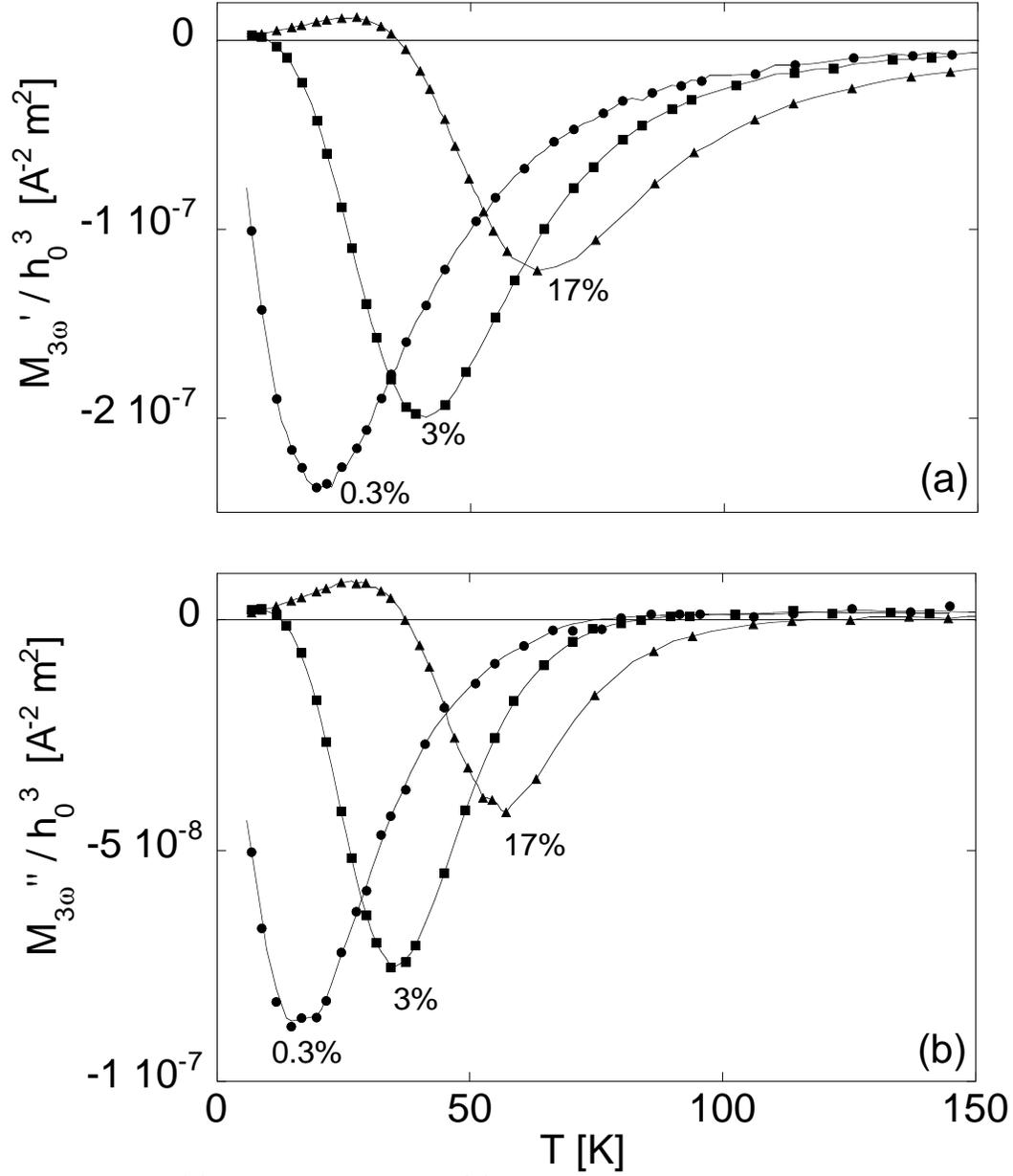,width=14truecm}}
\caption[]{The real (a) and the imaginary (b) parts of the cubic susceptibility vs. temperature measured at the angular frequency $3\omega$, with  $\omega/2\pi$ = 125 Hz, for the three samples with different volume concentration of magnetic particles.  $h_0$ = 200 A/m.
\label{susc3_H}
}
\end{figure}

Figures \ref{susc1_H} and \ref{susc3_H} clearly show that dipole-dipole interactions have a strong influence on the magnetic relaxation of the particle system.
Some recent experimental studies show that strongly interacting particle systems approach a spin glass phase at low temperatures. 
For example, spin glass characteristics such as: aging effects, \cite{Jonsson3} critical slowing down, \cite{Djurberg} and a divergence of the cubic equilibrium susceptibility \cite{Jonsson4} have been reported in such systems. 

In a model often used to analyze the dynamics of interacting particle systems, \cite{DoFiTr} the effect of interparticle interactions is accounted for by shifting the energy barrier distribution to higher  energies and thereby increasing the relaxation times. 
Some features shown in Figs. \ref{susc1_H} and \ref{susc3_H}, such as a shift of the maxima in $\chi(T,\omega)$ to higher temperatures and the appearance of weak positive peaks in $\chi_{3}(T,\omega)$ at low temperatures, can in principle be explained by a significant increase of the individual particle relaxation times. 
However, for the here investigated particle system, it has previously been shown that the low temperature magnetic relaxation for the most interacting sample resembles that observed for archetypal spin glasses. \cite{Jonsson3}
At temperatures below 45~K, the relaxation time spectrum broadens significantly and the magnetic relaxation is significantly different from that of the noninteracting sample. 
This together with the observed aging effect \cite{Jonsson2} give strong indications that the low temperature magnetic relaxation is dominated by collective particle dynamics. 
With increasing temperature, however, the time scale of the collective dynamics is gradually shifted to shorter time scales and the slow magnetic relaxation remaining is due to single-particle relaxation of a comparably small number of large particles. 
Still, the relaxation of these large particles is influenced by small particles surrounding them - relaxing particles will experience a magnetic field from neighboring polarized superparamagnetic particles, increasing the energy barriers and thereby the relaxation times. 
This gradual change of the magnetic relaxation with increasing temperature, from collective to single particle dynamics, obstructs the observation of critical slowing down in this particle system. 
For the same reason, it will not be possible to experimentally observe a divergent cubic equilibrium susceptibility. 
Nevertheless, the observation of collective behavior of the magnetic relaxation at low temperature makes it interesting to compare the behavior of the most interacting sample to that of a typical spin glass with long range interactions.
 
\begin{figure}[htb]
\centerline{\epsfig{figure=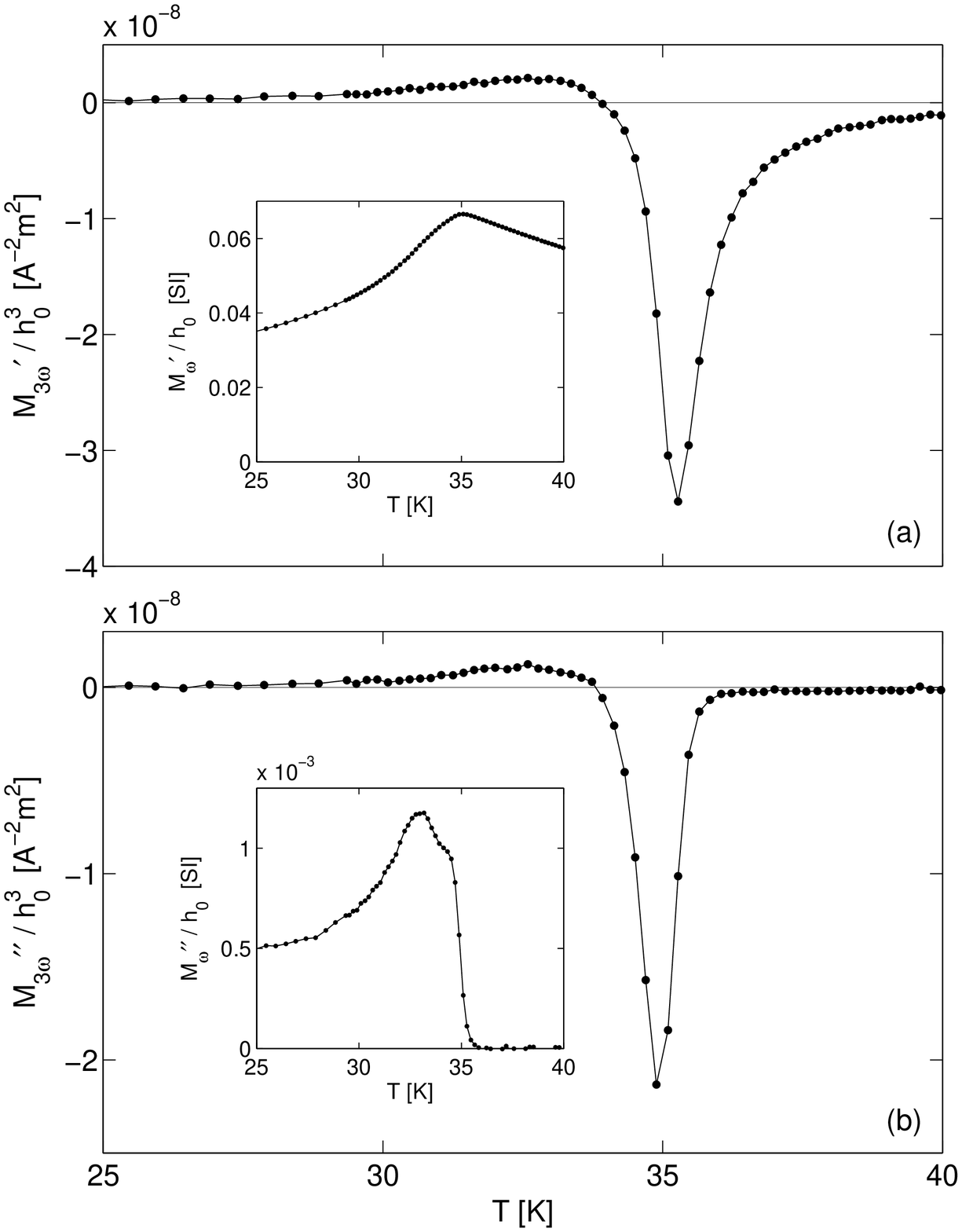,width=14truecm}}
\caption[]{The real (a) and the imaginary (b) parts of the cubic dynamic susceptibility vs. temperature measured at the angular frequency $3\omega$ for the Ag(11 at.\% Mn) spin glass sample. The linear susceptibility measured at $\omega/2\pi=125$~Hz is shown in the insets. $h_0$ = 1600~A/m.
\label{AgMn}
}
\end{figure}

Figure \ref{AgMn} shows the susceptibility of a Ag(11 at.\% Mn) spin glass measured at the frequency $\omega/2\pi = 125$ Hz and with an ac field amplitude of 1600 A/m. 
The real part  of the linear susceptibility has a sharp cusp at about the same temperature at which there is a sharp rise from zero of the imaginary part.
The two components of the cubic susceptibility have sharp negative peaks at high temperatures followed by broad positive peaks at low temperatures.
Before comparing this behavior with that of the most concentrated particle system, we will point out some fundamental differences between an interacting particle system and a spin glass.

In a particle system the single particle relaxation time has an exponential temperature dependence and in addition depends upon the particle volume and the anisotropy constant, while in a spin glass the individual spins have a relaxation time in the order of $10^{-12}$~s to $10^{-14}$~s. 
In a spin glass the slow dynamics is only due to collective phenomena and thus the imaginary component  of the susceptibility disappears when the maximum relaxation time of the spin system becomes shorter than the observation time $1/\omega$. 
The dynamics of the investigated particle system is, as discussed above, dominated by collective particle behavior at low temperatures, while at higher temperatures the slow dynamics are dominated by single particle relaxation. 
When comparing the two systems, we are thus restricted to the low temperature regime where both systems display collective dynamics. 
For the investigated interacting particle system, this implies temperatures $T < 45$~K.\cite{Jonsson3}

The linear and cubic susceptibilities of the most concentrated particle sample (see Figs. \ref{susc1_H} and \ref{susc3_H}) display features much broader in temperature as compared to the corresponding features observed for the spin glass sample. 
This can be explained as a combination of two effects: the polydispersivity of the particle system and that the maxima in the linear susceptibility as well as the negative peaks in the cubic susceptibility are partly due to single particle relaxation. 
It is tempting though to attribute the low temperature positive peaks in the interacting particle sample to collective dynamics, since also the spin glass shows broad positive peaks in the cubic dynamic susceptibility at low temperatures. 
This feature is found at temperatures $T < 40$ K, i.e., in the temperature range where the aforementioned study of the magnetic relaxation of the most interacting sample revealed evidence for collective particle behavior. 
%
%
                              
\section{Conclusions}
\label{Conclusions}

We have studied the linear and cubic dynamic susceptibilities of solid dispersions of maghemite nanoparticles with various strengths of the inter-particle interactions.
We have found that the expressions for the dynamical susceptibilities proposed by Raikher and Stepanov, describe the experimental results for the most dilute sample with a reasonable degree of accuracy.
Nevertheless, further developments of the theoretical modeling (including other symmetries of the anisotropy energy, effects of the damping parameter, etc.) would be desirable to reduce the gap between theory and experiments.

Concerning the features observed in the most concentrated sample an explanation in terms of single particle dynamics with equilibrium susceptibilities and energy barriers modified by the interactions cannot be excluded only on the basis of the present study.
However, previous evidence of collective magnetic behavior in the same system strongly suggests this as a more consistent interpretation of the obtained results.
This is also supported by the comparison of the results obtained for the most
concentrated particle system with the nonlinear response of an archetypal
spin glass.

\acknowledgements
This work was financially supported by The Swedish Natural Science Research Council (NFR).

\begin{table}[bht]
\caption[]{Parameters describing the particle system obtained by simultaneous fitting to the measured linear and cubic susceptibilities.}
\label{para}
\begin{tabular}{c c c c c} 
\emph{Volume distribution}& 
$V_0$ [nm$^3$]& & Anisotropy constant & $\tau_0$ [s] \\ \hline
Log-normal & 230 & s = 0.68 & $K$ = $1.4\times 10^4$ Jm$^{-3}$ &$5.7 \times 10^{-11}$ \\
Log-normal & 200 & s = 0.78 & $K_S$ = $2.3\times10^{-5}$ Jm$^{-2}$ &$1.1 \times 10^{-11}$ \\ 
Gamma & 210 & $\beta$ = 0.34 & $K$ = $1.2\times 10^4$ Jm$^{-3}$ &$2.1\times 10^{-10}$ \\ 
\end{tabular}
\end{table}


\begin{references}
\bibitem[*]{newadd}
{\em New address:} ABB Corporate Research, SE - 721 78 V{\"a}ster{\aa}s, Sweden.
\bibitem[\dag]{oldadd}
{\em On leave from:} Instituto de Ciencia de
Materiales de Arag\'{o}n, Consejo Superior de Investigaciones
Cient\'{\i}ficas-Universidad de Zaragoza, 50015 Zaragoza, Spain.
\bibitem{Neel}L. N\'{e}el, Ann. G\'{e}ophys. {\bf 5}, 99 (1949).
\bibitem{Brown}W. F. Brown, Phys. Rev. {\bf 130,} 1677 (1963).
\bibitem{Coffey}
W. T. Coffey, D. S. F. Crothers, Yu. P. Kalmykov, and E. S. Massawe, Phys. Rev. E {\bf 49}, 1869 (1994); 
W. T. Coffey, D. S. F. Crothers, Yu. P. Kalmykov, E. S. Massawe, and J. T. Waldron, J. Magn. Magn. Mater. {\bf 127}, L254 (1993);
$ibid$. {\bf 131}, L301 (1994).
\bibitem{Raikher1}
Yu. L. Raikher and V. I. Stepanov, Phys. Rev. B {\bf 52}, 3493 (1995).
\bibitem{Shliomis}
Yu. L. Raikher and M. I. Shliomis, Sov. Phys. JETP {\bf 40}, 526 (1974); 
M. I. Shliomis and V. I. Stepanov, J. Magn. Magn. Mater. {\bf 122}, 176 (1993).
\bibitem{Wernsdorfer}
W. Wernsdorfer et al., Phys. Rev. Lett. {\bf 78}, 1791 (1997). 

\bibitem{Jonsson1}
T. Jonsson, J. Mattsson, P. Nordblad, and P. Svedlindh, J. Magn. Magn. Mater. {\bf 168}, 269 (1997); 
P. Svedlindh, T. Jonsson, and J. L. Garc\'{\i}a-Palacios, $ibid$. {\bf 169}, 323 (1997).
\bibitem{Jonsson2}
T. Jonsson, J. Mattsson, C. Djurberg, F. A. Kahn, P. Nordblad, and P. Svedlindh, Phys. Rev. Lett. {\bf 75}, 4138 (1995).
\bibitem{Jonsson3}
T. Jonsson, P. Nordblad, and P. Svedlindh, Phys. Rev. B {\bf 57}, 497 (1998).
\bibitem{Djurberg}C. Djurberg, P. Svedlindh, P. Nordblad, M. F. Hansen, F. B\o dker, and S. M\o rup, Phys. Rev. Lett. {\bf 79}, 5154 (1997).
\bibitem{Bitoh}T. Bitoh, K. Ohba, M. Takamatsu, T. Shirane, and S. Chikazawa, 
J. Phys. Soc. Jpn. {\bf 62}, 2583 (1993); 
$ibid$. {\bf 64}, 1311 (1995);
J. Magn. Magn. Mater. {\bf 154}, 59 (1996).
\bibitem{Shirane}
T. Shirane, T. Moriya, T. Bitoh, A. Sawada, H. Aida, and S. Chikazawa, J. Phys. Soc. Jpn. {\bf 62}, 2583 (1993).
\bibitem{Garcia}J. L. Garc\'{\i}a-Palacios and F. J. L\'{a}zaro, Phys. Rev. B {\bf 55}, 1006 (1997).
\bibitem{Coffey3} W. T. coffey and B. V. Paranjape, Proc. R. Ir. Acad. Sect. A Math. Phys. Sci. {\bf 78}, 17 (1978).

\bibitem{Raikher2}Yu. L. Raikher and V. I. Stepanov, Phys. Rev. B {\bf 55}, 15005 (1997).
\bibitem{Klik} I. Klik and Y. D. Yao, J. Magn. Magn. Mater. {\bf 186}, 233 (1998).
\bibitem{Fiorani}D. Fiorani, J. Tholence, and J. L. Dormann,
J. Phys. C
{\bf 19}, 5495 (1986).
%
\bibitem{Jonsson4}
T. Jonsson, P. Svedlindh, and M. F. Hansen, Phys. Rev. Lett. {\bf 81}, 3976 (1998).
%

\bibitem{ferrofluids}
Ferrofluids Co., Nashua, New Hampshire.

\bibitem{Jiang}J. Z. Jiang, S. M{\o}rup, T. Jonsson and P. Svedlindh, Conf. Proc. Vol. 50, ICAME-95, Ed. I. Ortalli, SIF, Bologna (1996).

%
\bibitem{LakeShore}
Lake Shore, 575 McCorkle Boulevard, Westerville, OH 43082-8888 .

\bibitem{SQUID}
J. Magnusson, C. Djurberg, P. Granberg, and P. Nordblad, Rev. Sci. Instrum 
{\bf 68}, 3761 (1997).

%
\bibitem{Garcia2}
J. L. Garc\'{\i}a-Palacios and F. J. L\'azaro, Phys. Rev. B {\bf 58}, 14937 (1998).
\bibitem{Cregg} P. J. Cregg, D. S. F. Crothers and A. W. Wickstead, J. Appl. Phys {\bf 76}, 4900 (1994).

\bibitem{Shliomis2}M. I. Shliomis and V. I. Stepanov, Adv. Chem. Phys. {\bf 87}, 1 (1994).
\bibitem{Coffey2}W. T. Coffey, D. S. F. Crothers, J. L. Dormann, Yu. P. Kalmykov, E. C. Kennedy, and W. Wernsdorfer, Phys. Rev. Lett. {\bf 80}, 5655 (1998).

\bibitem{Gazeau}
F. Gazeau, J.C. Bacri, F. Gandron, R. Perzynski, Yu. L. Raikher and V.I. Stepanov, E. Dubois, J. Magn. Magn. Mater. {\bf 186}, 175 (1998).
\bibitem{Hasmonay}
E. Hasmonay, E. Dubois, J.-C. Bacri, R. Perzynski, Yu. L. Raikher and V.I. Stepanov, European Phys. J. B {\bf 5}, 859 (1998).


\bibitem{DoFiTr}J. L. Dormann, D. Fiorani, and E. Tronc, Adv. Chem. Phys. {\bf 98}, 283 (1997).

\end{references}
\end{document}